\documentstyle[12pt,epsf]{article}
\newcommand{\pslash}{p \!\!\!/}
\newcommand{\xslash}{x \!\!\!/}
\newcommand{\Dslash}{D \!\!\!/}
\newcommand{\nn}{\nonumber}
\newcommand{\beq}{\begin{eqnarray}}
\newcommand{\eeq}{\end{eqnarray}}

\newcommand{\pl}{Phys.Lett.\ }

\newcommand{\asl}{\alpha_s^{{}^{LATT}}}
\newcommand{\as}{\alpha_s^{{}^{\overline{MS}}}}
\begin{document}
\pagestyle{empty}\setcounter{page}{1}
\begin{flushright}
CERN-TH.7342/94 \\
LPTHE prep. 94/52 \\
ROME prep. 94/1022 \\
SHEP 94/95-03
\end{flushright}
\centerline{\bf{A GENERAL METHOD FOR NON-PERTURBATIVE}}
\centerline{\bf{RENORMALIZATION OF LATTICE OPERATORS}}
%%%%%%%%%%%%%%%%%%%%%%%%%%%%%%%%%%%%%%%%%%%%%%%%%%%%%%%%%%%%%%%%%%%%%
\vskip 0.8cm
\centerline{\bf{ G. Martinelli$^{a,b}$, C. Pittori$^{c}$,
C.T. Sachrajda$^d$, M. Testa$^a$ and A. Vladikas$^{e}$ }}
\centerline{$^a$ Dip. di Fisica,
Universit\`a degli Studi di Roma ``La Sapienza" and}
\centerline{INFN, Sezione di Roma, P.le A. Moro 2, 00185 Rome, Italy. }
\centerline{$^b$ Theory Division, CERN, 1211 Geneva 23, Switzerland.}
\centerline{$^c$ L.P.T.H.E., Universit\'e de Paris Sud, Centre d'Orsay,}
\centerline{91405 Orsay, France.}
\centerline{$^d$ Dep. of Physics, University of Southampton,}
\centerline{Southampton SO17 1BJ, U.K.
}
\centerline{$^e$ Dip. di Fisica, Universit\`a di Roma `` Tor Vergata'' and}
\centerline{INFN, Sezione di Roma II,}
\centerline{Via della Ricerca Scientifica 1, 00133 Rome, Italy.}
\begin{abstract}
We propose a non-perturbative method for computing the renormalization
constants of  generic
composite operators. This method is intended to reduce some systematic errors,
which are present when one tries to obtain physical predictions
 from the matrix
elements of lattice operators. We also present the results of a calculation
of the renormalization constants of several two-fermion operators, obtained,
with our method, by
numerical simulation of $QCD$, on a $16^3 \times 32$ lattice, at $\beta=6.0$.
The results of this simulation are encouraging, and further applications
to four-fermion operators and to the heavy quark effective theory are
proposed.
 \end{abstract}
\vskip 0.8cm
\begin{flushleft}
CERN-TH.7342/94 \\
November  1994
\end{flushleft}
\vfill\eject
\pagestyle{empty}\clearpage
\setcounter{page}{1}
\pagestyle{plain}
%\date{}
\newpage
%\end{titlepage}
\pagestyle{plain} \setcounter{page}{1}

\section{Introduction}
\label{sec:intro}
\par
Renormalization of lattice operators
is a necessary ingredient to obtain many physical
results from numerical simulations, such as meson decay constants,
form factors, structure functions, mixing amplitudes, etc.
In this paper we study the
renormalization of composite operators in lattice $QCD$.
We propose a method which avoids completely
 the use of lattice perturbation theory and allows a non-perturbative
determination
of the renormalization constants of any composite operator.
For illustrative purposes, we consider some specific applications
to matrix elements of two-fermion operators.
The approach is particularly useful in those cases,
such as the scalar or pseudoscalar densities,
where it is not possible to use the chiral Ward identities to determine
the renormalization constants non-perturbatively\footnote{In the following
we will denote the method to determine the renormalization constants, using
the Ward identities, as the Ward identity method.}  \cite{bo}--\cite{wi}.
Moreover,
our proposal can be applied to many other cases,
such as   the renormalization of
the four-fermion operators of the effective weak
Hamiltonian, and to heavy-light currents in the Heavy Quark Effective
Theory. Preliminary results have been presented in ref. \cite{dallas}.
A similar attempt, limited to the divergent part of the $\Delta I=1/2$
penguin-operator, can be found in ref. \cite{ber1}.
\par
Renormalized lattice operators must
correspond, in the limit of
infinite lattice cut-off, to finite chirally covariant operators, which
obey the same renormalization conditions as those in the continuum.
 With just a few exceptions, lattice perturbation theory is
 used to evaluate
the  renormalization constants of lattice operators.
The problem of mixing with lower dimensional
operators requires however, a non-perturbative
subtraction of all power divergences \cite{bo},\cite{te}--\cite{martsukuba}.
Apart from this special case,
the use of perturbation theory, in the computation of
multiplicative  constants and dimensionless mixing coefficients,
is well justified,
provided that the lattice spacing $a$ is sufficiently small,
i.e. $ a^{-1} \gg \Lambda_{QCD}$.
However, in those cases where the results from perturbation theory,
obtained using the bare lattice
coupling constant $\asl=g_{0}^{2}/4 \pi$ as an
expansion parameter, can be checked
using non-perturbative methods, there is generally a significant
discrepancy.
Several solutions to this problem have been proposed so far:
\begin{itemize}
\item
In order to improve the convergence
of the perturbative series,
the authors of ref. \cite{lepage} have  proposed a
 ``tadpole improved'' perturbation theory.
Nevertheless, ignorance of higher order contributions still  represents an
important source
of uncertainty in the extraction of physical results.
\item
In some limited cases, corresponding to finite
operators, namely the vector and axial vector currents,
and  the ratio of pseudoscalar and scalar densities, a fully
non-perturbative determination of the renormalization constants can be
obtained with the use of chiral Ward identities \cite{bo}--\cite{wi}.
\item
One may fix  non-perturbative renormalization conditions directly on
hadronic matrix elements. This procedure was used in ref.
\cite{gav1}, to subtract the divergent part of the $\Delta I=1/2$ operator.
The price is that one has to sacrifice
 a physical prediction, for any subtraction imposed
on hadronic states. Thus, when several
renormalization
conditions are necessary,
 one  looses much predictive power \cite{bo,te}.
\end{itemize}
\par
{\it
Our proposal is to impose
re\-nor\-ma\-li\-za\-tion con\-di\-tions
non-per\-tur\-ba\-ti\-vely,  di\-rectly on quark and
gluon Green functions, in a fixed gauge, with given off-shell
ex\-ter\-nal sta\-tes, with large vir\-tua\-li\-ties.}
\centerline{}
\centerline{}
\par
The method consists in  mimicking  what is usually done in perturbation
theory.
One fixes the renormalization conditions of a
certain operator by imposing that suitable Green functions, computed
between external off-shell quark and gluon states, in a fixed gauge,
coincide with their tree level value. For example, if we consider
the generic two-quark operator $O_\Gamma= \bar \psi \Gamma \psi$,
we may impose the condition
\beq Z_\Gamma \langle p \vert O_\Gamma \vert p \rangle\vert_{p^2=-\mu^2}=
\langle p \vert O_\Gamma \vert p \rangle_0, \eeq
where $\Gamma$ is one of the Dirac matrices and $
\langle p \vert O_\Gamma \vert p \rangle_0$ is the tree level matrix element.
The extension to more complicated cases, including four-fermion operators
and operator mixing, is straightforward.
This procedure defines the same renormalized operators, i.e. the same
Wilson coefficient functions, in all  regularization schemes,
 provided they are expressed in terms of the same renormalized
coupling constant. However, the coefficient functions now depend on
the external states and on the gauge, which must be specified.
 \par  In principle this scheme completely solves
the problem of  large corrections in lattice perturbation theory,
 which are automatically
included in the renormalization constants.
Matching of  lattice operators to
the corresponding ones, renormalized in a
continuum renormalization scheme (for definiteness, the $\overline{MS}$
scheme),
then requires continuum perturbation theory only.
When  next-to-leading corrections are known, the error
on the physical matrix elements is of  $O((\as)^2)$, in the continuum
perturbative expansion, which is expected to have a better convergence.
A continuum perturbative calculation of the matching conditions
 is a common step in  all approaches, standard or
 ``tadpole improved" perturbation theory and non-perturbative
renormalization.
Our proposal  can be applied to any composite operator,
unlike the  Ward identity method, which
 is limited to a few, albeit very important, cases\footnote
{It is still true that the statistical
accuracy is in general slightly better with the Ward identities
method, in those cases
where it can be applied, see sec. \ref{sec:nonper}.}.
\par  \par The feasibility of using Green functions computed between
 quark states,
in a fixed gauge, was already discussed in ref. \cite{wi}, in the context
of non-perturbative renormalization, using Ward identities. Indeed, in the case
of the axial current, it was shown
that the signal was rather good and the result was in agreement with the
corresponding hadronic gauge invariant determination.
\par As far as the renormalization conditions are concerned,
our proposal is expected to work, whenever it is  possible to
fix the virtuality of the external states $\mu$ and to satisfy
the condition $ \Lambda_{QCD} \ll \mu \ll 1/a $, in order
to keep under control both non-perturbative and discretization effects.
We stress that this  requirement is common to all methods.
The results of our numerical investigation
are encouraging: they suggest that there does exist a ``window" in $\mu$,
where the method can be applied,
at values of the lattice coupling constant used in
current lattice simulations.
\par The plan for the remainder of the paper is as follows. In the next
section the basic formulae necessary to define our non-perturbative
method are presented, using as an example logarithmically divergent
two-quark operators, such as the scalar and pseudoscalar densities. The
discussion is generalised to all composite operators in section 3, and
in the following section the implementation of the method in numerical
simulations for two-quark operators is discussed. In section 5 we give
some details about the perturbative evaluation of the renormalisation
constants on lattices of finite volume. The values of the
renormalisation constants obtained in a numerical simulation are
presented in section 6, and are compared to the results from
perturbation theory. The final section contains our conclusions, a
summary of the numerical results and a discussion of possible further
applications of the ideas presented in this paper.
 \section{Non-perturbative renormalization conditions}
\label{sec:idea}
\par
In this section, the non-perturbative renormalization scheme
is presented in detail.  We also
discuss the conditions  which must be satisfied for our method to be
applicable. \par
To facilitate the discussion, it is convenient to classify composite operators
into three main classes, according to their ultra-violet behaviour, as
$a \to 0$:
\begin{enumerate}
\item {\bf Finite operators}: these include the vector and axial vector
currents. Another example  is  the ratio
of the  scalar and pseudoscalar renormalization constants, $Z_S/Z_P$. For
these cases, the Ward identity method is also applicable.
\item {\bf Logarithmically divergent operators}:
this is a large family of operators, relevant to hadron phenomenology.
It includes
   scalar and pseudoscalar densities,  four-fermion $\Delta F=2$
 operators, some components of the energy-momentum tensor, etc.
\item
 {\bf Power divergent operators}: these operators are present
when  mixing with lower dimensional
operators is possible. This happens in regularizations which have  an intrinsic
mass scale, such as  the lattice regularization.
Examples are often encountered in phenomenological applications of
lattice $QCD$, such
as  four-fermion operators relevant to
$\Delta I=1/2$ transitions or operators of order $1/m$
in the heavy quark effective theory (HQET) \cite{effi}--\cite{efff}.
\end{enumerate}
 \par For simplicity,
we first consider the two-fermion operators of class 2., and extend
the discussion to other operators in this class, and also to those
in classes
 1. and 3., in the next section.
Thus, the formulae given below apply directly to the  pseudoscalar
and scalar densities. Throughout our discussion, we assume that discretization
errors are negligible: in field theory language, this is equivalent
to the statement that
the renormalized Green functions do not differ appreciably from their
values in the limit of an infinite  ultra-violet cut-off.
The discussion is presented in the  limit of small quark masses and
in  Euclidean space-time. The discretisation of the quark action is assumed
to be performed {\it \`a la} Wilson, characterized by   explicit chiral
symmetry
breaking of the Lagrangian\footnote{This includes the SW-Clover action
\cite{slami}.}. The extension to staggered fermions
 is straightforward.
\par Let us consider the forward amputated Green function  $\Gamma_{O}(pa)$,
of a two-fermion bare lattice  operator $O(a)$, computed between off-shell
quark
states  with four-momentum $p$, with $p^2=\mu^2$, and
in a fixed gauge, for example the Landau gauge.
Without gauge-fixing all Green functions
computed between quark and gluon external
states are zero. This is  also true when the gauge is not
completely fixed, e.g. in the Coulomb gauge.
We define the renormalized operator $O(\mu)$, by introducing
the renormalization constant $Z_O$
\beq O(\mu)= Z_O\Bigl(\mu a, g(a)\Bigr) O(a). \label{eq:zren} \eeq
$Z_O$ is found, by imposing the renormalization condition
\beq   Z_O\Bigl(\mu a,  g(a)\Bigr) Z^{-1}_\psi
\Bigl(\mu a,  g(a)\Bigr)
\Gamma_{O}(pa)\vert_{p^2=\mu^2}=1, \label{eq:zdef} \eeq
where $Z_\psi$ is the field renormalization constant, to be defined below.
This  procedure defines a renormalized operator $O(\mu)$
which is  independent of the regularization scheme
\cite{altarelli}--\cite{ciu}.
 It depends,  however, on the external states and on the gauge. This does not
affect the final results, which, combined with the continuum calculation
of the renormalization conditions, at any given order of perturbation
theory,
will be gauge invariant
and independent of the external states.
Let us specify the different quantities entering  eq. (\ref{eq:zdef}).
$\Gamma_{O}$ is defined in terms of the expectation value\footnote{
 ``Expectation value" means, as  usual, that one averages the Green functions
 over the gauge field configurations, generated by  Monte Carlo
simulation, see sec. \ref{sec:duef}.}  of
the non-amputated Green function $G_{O}(pa)$, and
of the quark propagator $S(pa)$
\beq \Gamma_{O}(pa)= \frac{1}{12} {\rm tr} \left( \Lambda_{O}(pa)\hat  P_O
\right), \label{eq:proj} \eeq
where
\beq \Lambda_{O}(pa)=  S(pa)^{-1} G_{O}(pa) S(pa)^{-1}. \label{eq:lam}
\eeq
$\hat P_O$ is a suitable projector on the tree-level operator: $\hat P_O=
\hat 1$ ($\hat P_O= \gamma_5$) for the scalar (pseudoscalar) density.
The factor $1/12$ ensures the correct overall normalization of the
trace (colour$\times$spin=12).
Projectors are very convenient when defining   Green functions,
particularly in the non-perturbative case. They
 have been extensively used in refs. \cite{buras,ciu}. Of course  one can
also use other definitions of $Z_O$.
\par
$Z^{1/2}_\psi$ is the renormalization constant of the fermion field.
It can be defined in different ways, some of which are equivalent
perturbatively. Beyond perturbation theory, the most natural
definition of $Z_\psi$ is obtained from the amputated Green function
of the conserved vector current $V^{C}$. Indeed, one knows
that for $V^{C}$ the renormalization constant is equal to one:
\beq
Z^{-1}_{V^{C}}= \Gamma_{V^C} \times Z^{-1}_{\psi}=
\frac{1}{48} {\rm tr}
\left( \Lambda_{V^{C}_{\mu}} (pa)\gamma_{\mu}\right)\vert_{p^2=\mu^2} \times
Z^{-1}_{\psi}=1,
\eeq
which implies
\beq
Z_{\psi}=\frac{1}{48} {\rm tr}
\left( \Lambda_{V^{C}_{\mu}} (pa)\gamma_{\mu}\right)\vert_{p^2=\mu^2}.
\label{eq:zpsi}
\eeq \par
Equations (\ref{eq:zdef})--(\ref{eq:zpsi}) completely define our method.
In the remainder of this section, we discuss some important aspects
concerning its applicability. \par In the above formulae, for simplicity,
we have always
considered only forward matrix elements. In general, one has the freedom
to define the renormalization conditions at different external
momenta $p$ and $p^\prime$ ($p \neq p^\prime$).
The virtualities of the  quark states must be  much larger than
$\Lambda_{QCD}$.
 The reason is that, in
order to obtain the physical result, we have to combine the matrix element
of the renormalized operator $O(\mu)$, with a Wilson coefficient function.
The latter is computed
in continuum perturbation theory, by expanding in
 $\as$ at a scale of order $\mu$. Thus,
for the validity of this perturbative calculation, $\mu$ must be large.
An important question in our program is whether it is possible to find, on
the lattice, a scale $\mu$ which is sufficiently low, in order to
have small  $O(a)$ effects,
and sufficiently large, in order to have small higher order corrections.
The range of $\mu$, where the above conditions are satisfied, depends
on the value of $g_0^2$ of the numerical simulation.
We stress that, if such a  window
does not exist, in current lattice simulations,
an accurate matching of  lattice operators to
 continuum ones becomes impossible,
not only in our approach, but also with any other method.
\par It might be expected that the condition $\mu \gg \Lambda_{QCD}$ ensures
that perturbation
theory is valid. When
spontaneous symmetry breaking occurs however, as  is the case in $QCD$, a large
value of $\mu$ may  not be enough, because of the
presence of  the Goldstone boson, the pion in our case.
At low momentum transfers $q=
p-p^\prime$,  Green functions  can receive  a
non-perturbative contribution from
the pion pole, which cannot be computed. The na\"\i ve expectation is that this
 effect vanishes
 as $1/q^2$, at large $q^2$. However, with fermion fields, this
contribution is proportional to  $1/p^2=1/\mu^2$, even when $q^2=0$.
The proof is given in the appendix.
We conclude that a large value of $\mu$ solves the problem. At low values
of $\mu^2$, however, for the pseudoscalar density
and the axial current, we expect to find visible  effects,
see secs. \ref{sec:casivari} and \ref{sec:nonper}.
\par
A  danger to the non-perturbative scheme may come from the
presence of  Gribov copies. We do not address  this problem here;
it  has been investigated in refs. \cite{altri,gribov}.
The results of
ref. \cite{gribov} indicate that,
at least for the quantities of interest in the present study,  the ``Gribov
uncertainty'', in current lattice simulations, is at most of the same order as
 the statistical error.
\par One may imagine applying the renormalization conditions at very
large values of $\beta$, on lattices which have small physical volumes.
This might seem to overcome the problem of the existence of the window in
$\mu$.
For  physical applications however, it is necessary to evolve the
renormalization constants to smaller values of $\beta$.
Non-perturbative contributions, which are not detected at large $\beta$'s,
may then become important. This demonstrates the necessity,
for any  method,
of the existence of a  window $\Lambda_{QCD} \ll \mu \ll 1/a$, where
$a$ is the lattice spacing used in  numerical simulations of physical
quantities.  Only then are  non-perturbative effects and  lattice artefacts
small simultaneously.
\section{Applications of the method to generic composite operators}
\label{sec:casivari}
In the previous section, using the renormalization of the scalar and
pseudoscalar densities as a prototype,
the general strategy of the non-perturbative scheme has been presented.
 In the present section the necessary modifications
needed in other cases are briefly described. We will follow the
classification of operators   introduced in sec. \ref{sec:idea}.
\subsection{Finite operators}
 In the case of vector and axial vector
currents, we do not have the freedom of  fixing the
renormalization conditions in an arbitrary way:
the renormalization conditions are acceptable only
when  they are compatible with the Ward identities.
To be specific, one can apply  eq. (\ref{eq:zdef}) to
 the local vector current in order
to determine $Z_{V^L}$
\beq Z_{V^L} = \frac{Z_\psi}{\Gamma_{V^L}}, \label{eq:zvl} \eeq
We now show  that this definition of $Z_V^L$,
which uses  the  general prescription given
in the previous section,  coincides
with the one derived from the vector current Ward identity  on quark
states \cite{bo,ka}\footnote{For the remainder of this subsection we work
in lattice units, setting $a=1$.}
\beq
q_{\mu} \Bigl[Z_{V^L} \Lambda_{V^L_\mu}(p+\frac{q}{2},p-\frac{q}{2})
\Bigr]=-i\Bigl( S^{-1}(p+\frac{q}{2}) - S^{-1}(p-\frac{q}{2})\Bigr).
\label{eq:pallino}\eeq
where $\Lambda_{V^L_\mu}(p+q/2,p-q/2)$ is the amputated
vertex with momentum transfer $q$.
By applying $\partial / \partial q_\rho$, at $q=0$, on both sides of eq.
(\ref{eq:pallino}), one gets
\beq
Z_{V^L}\Bigl( \Lambda_{V^L_\rho}(p) + q_\mu \frac{\partial}{\partial q_\rho}
\Lambda_{V^L_\mu}(p+\frac{q}{2},p-\frac{q}{2})\vert_{q=0}\Bigr)
=-i \frac{\partial}{\partial p_\rho}S^{-1}(p).
\label{eq:pinco} \eeq
The second term on the l.h.s. vanishes when $q \to 0$.
By tracing eq. (\ref{eq:pinco}) with $\gamma_{\rho}$,
we then recover the definition of $Z_{V^L}$ of eq. (\ref{eq:zvl}).
\par One is tempted to follow the same path for the local axial vector current,
i.e. to define
\beq Z_{A^L} = \frac{Z_\psi}{\Gamma_{A^L}}.  \label{eq:zal} \eeq
However this definition of $Z_{A^L}$   is not equivalent to the one
obtained by the Ward identity
\beq
q_{\mu} \Bigl[Z_{A^L} \Lambda_{A^L_\mu}(p+\frac{q}{2},p-\frac{q}{2})
\Bigr]=-i\Bigl( \gamma_5 S^{-1}(p+\frac{q}{2}) + S^{-1}(p-\frac{q}{2})
\gamma_5 \Bigr).
\label{eq:awi}\eeq
To see this we proceed as for the vector current,  applying $\partial /\partial
q_\rho$
to both sides of eq.  (\ref{eq:awi}).
\beq Z_{A^L}\Bigl( \Lambda_{A^L_\rho}(p) + q_\mu
\frac{\partial}{\partial q_\rho}
\Lambda_{A^L_\mu}(p+\frac{q}{2},p-\frac{q}{2})\vert_{q=0}\Bigr) =
 \nonumber \eeq
\beq  -\frac{i}{2}\Bigl( \gamma_5 \frac{\partial}{\partial p_\rho}S^{-1}(p)
-\frac{\partial}{\partial p_\rho}S^{-1}(p) \gamma_5 \Bigr).
\label{eq:dawi} \eeq
 In the axial case, however,  the second term
 on the l.h.s. does not vanish, as $q \to 0$ and, indeed,
this term is necessary to saturate
the Ward identity, in presence of a massless Goldstone boson.
As demonstrated in the appendix however, this term becomes negligible
for high values of $p^2$. In this case, by tracing eq. (\ref{eq:dawi}),
one recovers (\ref{eq:zal}). In conclusion,
we expect to find a value of
$Z_{A^L}$ close to those obtained from the Ward identity on hadron or
 on quark states \cite{wi}, by using eq. (\ref{eq:zal}), at large
$\mu^2$.
\par  $Z_S/Z_P$ is another finite quantity, which is completely fixed by
the Ward identities, even though the renormalization conditions
on $Z_S$ and $Z_P$  are separately arbitrary. In other words
one is free to  decide the renormalization conditions
of one of the two operators,
the scalar density  say, and the value of $Z_P$ will automatically follow.
The procedure of sec. \ref{sec:idea}, respects the Ward identities
and thus  guarantees that the correct value of $Z_S/Z_P$ is obtained.
This is clearly true only at large values of $\mu^2$,
for the same reasons as for
the axial vector current.
\subsection{Logarithmically divergent operators}
For this category of operator only the question of operator-mixing is left to
be discussed.
There are two kinds of mixing: the one which can occur also in
the continuum and the one induced by the explicit lattice
chiral symmetry breaking.
We are only  concerned  here with the latter case,
and hence restrict our discussion to operators
which renormalize multiplicatively in the continuum.
To illustrate our arguments, we consider  the four-fermion operator
\beq O^{\Delta S=2} = \Bigl(\bar s \gamma^\mu (1-\gamma_5) d \Bigr)
\Bigl(\bar s \gamma_\mu (1-\gamma_5) d \Bigr), \label{eq:ds2b} \eeq
which is relevant to the calculation of the $K^0$--$\bar K^0$ mixing
amplitude.
Because of the Wilson term, this operator mixes with operators of different
chiralities \cite{bo,te,4ferm}
\beq  O^{\Delta S=2}_{{\rm cont}} = Z_O \Bigl( O^{\Delta S=2}_{{\rm
latt}} + \sum_i Z_i O^i \Bigr),\label{eq:omix} \eeq
where the operators $O^i$ are given by $O^1= (\bar s d )(\bar s d),
O^2=(\bar s
\gamma_5 d)(\bar s \gamma_5 d)$, etc. \cite{4ferm}.
 The mixing coefficients $Z_i$ are finite
functions of $g_0^2(a)$.
The over-all renormalization $Z_O$ is necessary to eliminate
the logarithmic divergence.
\par We now sketch the renormalization procedure for $O^{\Delta S=2}$.
Let us start from the four-point amputated, Green function of
$O^{\Delta S=2}_{{\rm latt}}$, between quark states with momentum $p^2=\mu^2$
\beq \Lambda_{\alpha \beta \gamma \delta}^{ABCD}(pa),
\label{eq:g4} \eeq
where $\alpha, \beta \dots$ and $A, B \dots$ are spin and colour
indices respectively.
To find the mixing coefficients $Z_i$, one uses suitable projectors
$\hat P$, as
was done for two-quark operators\footnote{ In order to simplify the
presentation,  we do not discuss here the
complications arising from the different possible colour structures.},
for example
\beq {\rm tr } ( \Lambda \hat P )= \Lambda_{ \alpha \beta
\gamma \delta }^{AABB}( \gamma_5)_{ \beta \alpha}
( \gamma_5 )_{ \delta \gamma}. \eeq
By projecting on all the possible colour-spin structures, corresponding
to the operators $O^i$, one obtains a system of linear equations in
the mixing coefficients $Z_i$.  Using the $Z_i$, determined in this way,
one can compute the Green function of the subtracted operator
$O^s=O^{\Delta S=2}_{{\rm latt}}+ \sum_i Z_i O^i$. From the amputated vertex of
$O^s$, one can finally remove the logarithmic divergence, by imposing the
renormalization condition
\beq Z_O \Bigl( \mu , g(a) \Bigr) Z^{-2}_{ \psi }
\Bigl( \mu , g(a) \Bigr) \Gamma^s_O ( pa) \vert_{p^2 =
\mu^2} = 1, \eeq
where
\beq \Gamma^s_O \sim { \rm tr } (
\Lambda^s \hat P_{ L \times L} ) ,\eeq
and $P_{ L \times L}$ is the suitable projector on the
$ \gamma_ \mu ( 1 - \gamma_5 ) \otimes \gamma^\mu (
1 - \gamma_5 ) $ Dirac tensor.
\subsection{Power divergent operators}
\begin{figure}
\centering
\epsfysize=2.5cm
\epsfxsize=7.0cm
\epsffile{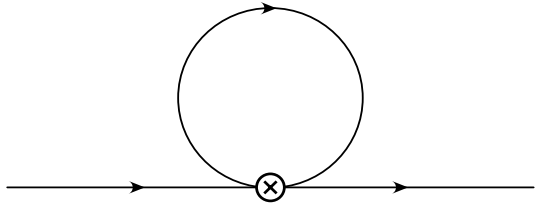}
\caption{Diagram which mixes the four fermion operator in
eq. (\ref{eq:dt12}) with the scalar
and pseudoscalar densities. The mixing is induced by the chiral
breaking term present in the Wilson and Clover actions.}
\label{fig:penguin}
\end{figure}

Composite operators are divergent in $1/a$  when they can mix with lower
dimensional operators. Examples are given by the
 operators relevant to deep inelastic scattering and by
the penguin-operators  appearing in the $\Delta S=1/2$ Hamiltonian.
Let us consider the $\Delta S=1/2$ operator
\beq O^{ \Delta S=1/2}= ( \bar s \gamma_\mu ( 1 - \gamma_5 )
d ) \sum_i (\bar q_i \gamma^\mu ( 1 + \gamma_5 ) q_i) .\label{eq:dt12} \eeq
$ O^{ \Delta S=1/2}_{{\rm latt}}$ mixes with the scalar and pseudoscalar
denisties, even at zeroth order in $\alpha_s$, through the diagram in
fig. \ref{fig:penguin}\footnote{ $O^{\Delta S=1/2}_{{\rm latt}}$ also  mixes
 with the chromo-magnetic operator, e.g. $\bar s \sigma_{\mu \nu}
G^{\mu \nu} d$, which is not being considered here.}
\beq O^{\Delta S=1/2}_{{\rm cont}} =
Z_O O^{\Delta S=1/2}_{{\rm latt}}+ Z_1 (\bar s d )+ Z_2 (\bar s  \gamma_5
d )+ \dots. \label{eq:ds12} \eeq
To find the coefficients $Z_{1,2}$, one imposes that the two-point
Green function of the operator on the r.h.s. of eq.
(\ref{eq:ds12}), on off-shell quark states,  is zero \cite{ber1}.
Once  all the divergent parts have been removed, one can impose the
renormalization conditions, necessary to remove the over-all logarithmic
divergence, as discussed above.
 \section{Implementation of the method in numerical simulations}
\label{sec:duef}
\par
To implement the ideas of the previous section,
let us consider the generic local two-fermion operator
\beq O_{\Gamma}(x) = \bar{\psi}(x)\Gamma \psi(x) \label{eq:2ferm} \eeq
where $\Gamma$ is any Dirac matrix.
The lattice, bare Green function of $O_{\Gamma}$, between off-shell quark
states,
can be obtained from
\beq
G_{O}( x, y )=\langle \psi(x) O_{\Gamma}(0)\bar{\psi}(y) \rangle=
\frac{1}{N}\sum^{N}_{i=1} S_i(x \vert 0) \Gamma S_i(0 \vert y),
\label{eq:gxy} \eeq
where $i=1,N$ labels the gauge field configurations and
$S_i(x \vert 0) $ is the quark propagator on the single configuration $i$,
obtained by inverting the discretized Dirac operator. We denote by
 $S_i(x \vert y)$ the inverse Dirac operator, which is not translationally
invariant, in contrast to $S(x-y)$, which is the usual quark propagator.
    The gauge field configurations
are generated with some standard, gauge-invariant algorithm;
the gauge fields and the quark
propagators are then rotated into the Landau gauge, which
 will be  defined in sec. \ref{sec:nonper}.
 From eq. (\ref{eq:gxy}), using $S_i(x \vert y)= \gamma_5
S_i(y \vert x )^\dagger \gamma_5$, one gets
\beq
G_{O}(pa)&\equiv &\int d^4 x d^4 y  e^{-i \,( p \cdot x- p \cdot y)}
G_O( x ,  y ) \nn \\
&=&\frac{1}{N} \sum_{i=1}^{N} \Bigl(\int d^4 x S_i(x \vert 0)
e^{-i \, p \cdot x} \Bigr) \Gamma \Bigl(\int d^4 y S_i(0 \vert y)
e^{i \, p \cdot y} \Bigr)
\nn \\ &=&\frac{1}{N} \sum_{i=1}^{N}  S_i(p \vert 0)
\Gamma  \gamma_5 \Bigl(\int d^4 y S^\dagger_i(y \vert 0)
e^{i \, p \cdot y} \Bigr) \gamma_5
\nn \\& =&\frac{1}{N} \sum_{i=1}^{N} S_i(p \vert 0)
 \Gamma  \Bigl( \gamma_5 S^\dagger_i(p \vert  0) \gamma_5
 \Bigr) ,
\label{eq:nuovo}
\eeq
where we have defined
\beq
S_i(p \vert y)=\int d^{4}x S_i(x \vert y)
e^{-i \,( p \cdot x)},
\eeq
because, on a single configuration, the quark propagator
is not translationally invariant. Traslational invariance is recovered
after averaging over the configurations. We can then write
\beq
S(pa)= \langle S(p \vert 0) \rangle= \frac{1}{N}\sum^{N}_{i=1} S_i(p \vert 0)
. \label{eq:nuovo1} \eeq
\par
Eqs. (\ref{eq:nuovo}) and (\ref{eq:nuovo1}) give $G_O(pa)$ and
$S(pa)$ in terms of quark propagators, computed by inverting  the Dirac
operator in the  presence of a background gauge field configuration, generated
by numerical simulation. With these quantities at hand, one can
apply the strategy and  the  formulae,
introduced in sec. \ref{sec:idea}. We add some information, that can help
the reader in practical applications.
The quark wave-function renormalization constant $Z_\psi$ has been
 computed using
\beq
Z_{\psi}= \Gamma_{V^L} \times Z_{V^L}= \frac{1}{48} {\rm tr}
\left( \Lambda_{V^L_{\mu}} (p a)\gamma_\mu\right)\vert_{p^2=\mu^2}\times
  Z_{V^{L}}
\label{eq:zpsil}
\eeq
where $\Lambda_{V_\mu^L}(pa) $ is defined as in eq. (\ref{eq:lam});
the trace
is taken over colours and spins and $Z_{V^L}$ is the renormalization
constant of the local vector current $V^L_\mu(x)=\bar \psi(x)
\gamma_\mu \psi(x)$. Equations (\ref{eq:zpsi})
and (\ref{eq:zpsil}) are two equivalent definitions of
$Z_\psi$. We have used eq. (\ref{eq:zpsil}) instead
of eq. (\ref{eq:zpsi}), because the former is more convenient when,
as in our case, one has only computed  quark propagators which
stem from the  origin. The  three-point ($q-V-q$) Green function is obtained by
inserting
the local  vector   current in the origin.    $Z_{V^L}$ can be obtained with
very high
accuracy from ratios  of matrix elements of the conserved and
local currents \cite{ma,msv}.
{}From the amputated Green functions
\beq \Gamma_{1}(pa)=\frac{1}{12} {\rm tr} \Bigl( \Lambda_1(pa)\hat 1 \Bigr) ,
\nn \eeq
\beq \Gamma_{\gamma_5}(pa)=\frac{1}{12} {\rm tr} \Bigl( \Lambda_
{\gamma_5}(pa) \gamma_5 \Bigr) , \nn
\eeq
\beq \Gamma_{\gamma_\mu \gamma_5}(pa)=\frac{1}{48} {\rm tr} \Bigl( \Lambda_
{\gamma_\mu \gamma_5}(pa) \gamma_5 \gamma_\mu \Bigr) ,
\eeq
 by imposing the renormalization conditions (\ref{eq:zdef}),
at several
values of $p^2=\mu^2$, to be specified in sec. \ref{sec:nonper},
we have
determined $Z_{S,P,A}\Bigl(\mu,g(a)\Bigr)$,
as functions of $\mu^2$. \par If we wish to compute
  $Z_{V^L}$  using the procedure defined in sec. \ref{sec:casivari}
we clearly cannot take $Z_\psi$ from eq. (\ref{eq:zpsil}).
{}From the Ward identity we have
\beq Z_\psi = - i \frac{1}{12}
{\rm tr} \Bigl( \frac{\partial S(p)^{-1}}{\partial \pslash}
\Bigr)\vert_{p^2=\mu^2} \label{eq:zpsiq}
\eeq
where $\partial S(p)^{-1}/\partial \pslash = 1/4 \times
\gamma_\rho \partial S(p)^{-1}/\partial p_\rho$.
To avoid derivatives with respect to a discrete variable,
we have used
\beq Z_\psi^{\prime}\Bigl(\mu , g(a) \Bigr)= \frac {{\rm tr } \Bigl(
-i \sum_{\lambda=1,4} \gamma_\lambda \sin( p_\lambda a ) S^{-1}(pa)
\Bigr)}{4 \sum _{\lambda=1,4} \sin^2 p_\lambda a }\vert_{p^2=\mu^2}.
\label{eq:zpsi1} \eeq
In one loop perturbation theory, $Z_\psi^{\prime}=Z_\psi$,
in the Landau gauge\footnote{In general, they  differ  by a finite
term of order $\alpha_s$.}. By using
\beq \Gamma_{\gamma_\mu }(pa)=\frac{1}{48} {\rm tr} \Bigl( \Lambda_
{\gamma_\mu }(pa) \gamma_\mu \Bigr) , \label{eq:gvl}
\eeq
and imposing the renormalization conditions (\ref{eq:zdef}), we have determined
$Z_{V^L}$ at several $\mu^2$. From the general discussion of the previous
section, one expects that $Z_{V^L}$, defined through the present procedure,
 is independent of $\mu^2$, up to
terms of $O(a)$, as  has been effectively found,
see sec. \ref{sec:nonper}.

In actual numerical simulations, the  renormalization procedure
could be obscured by lattice artefacts, i.e.  terms of $O(\mu a)$, which
are present
 when one imposes
the renormalization conditions at large values of $\mu^2$,  required  to avoid
large
non-perturbative or higher  order effects.
 As shown in ref.\cite{noi} (we use here the same notation),  in order to
reduce discretization errors, due to the finite value of the
lattice spacing, one can compute correlation functions using a
nearest-neighbour
improved fermion action \cite{slami} \beq
S^{I}_{F}=
S^{W}_{F} +
a^{4} \sum_{x,\mu \nu} \left[-ig_{0}
\frac{ar}{4}\bar{\psi}(x)
\sigma_{\mu \nu}F_{\mu \nu}(x)\psi(x) \right],
\label{eq:si}
\eeq
where $S^{W}_{F}$ is the usual Wilson action, $r$ is the Wilson parameter
and $F_{\mu\nu}$ is the lattice field strength tensor.
Moreover,
it is also necessary to use improved operators
$O_{\Gamma}^{I}$ defined as
\beq
O_{\Gamma}^I(x) =
\bar{\psi}(x) \overleftarrow{R}_{x}
\Gamma
\overrightarrow{R}_{x} \psi(x),
\label{eq:oi}
\eeq
where
the rotations of the fermion fields
are defined as
\beq
\overrightarrow{R}_{x}=
\Bigl(1-a {r\over{4}} (\overrightarrow{\Dslash}_{x} - m_0)\Bigr)
{}~~~,~~~
\overleftarrow{R}_{y}=
\Bigl(1+a {r\over{4}} (\overleftarrow{\Dslash}_{y} + m_0)\Bigr).
\label{eq:rotaz}
\eeq
By using the improved action and operators, one  eliminates discretization
errors of $O(a)$, and only those of  $O(\alpha_s a)$ or $O(a^2)$, or higher,
remain.
In eq. (\ref{eq:oi}), $\overrightarrow R_x$ and
 $\overleftarrow R_y$ can be replaced by
\beq
\overrightarrow{R^\prime}_{x}=
1-a {r\over{2}} \overrightarrow{\Dslash}_{x}
{}~~~,~~~
\overleftarrow{R^\prime}_{y}=
1+a {r\over{2}} \overleftarrow{\Dslash}_{y},
\label{eq:rotaz1}
\eeq
in computations of  the hadron spectrum and of on-shell
operator matrix  elements. As  discussed in refs.\cite{tass2,newi},
 the improvement procedure is equivalent to expressing all correlation
functions
in terms of  ``effective'' quark propagators
\beq
S^{eff}_i(x \vert 0)=
\Bigl(1-a{r\over{2}}
\overrightarrow{\Dslash}_{x} \Bigr)
S^{I}_i(x \vert 0)
\Bigl(1+a{r\over{2}}
\overleftarrow{\Dslash}_{y=0} \Bigr)
\label{eq:seff}
\eeq
where $S^{I}_i(x \vert 0)$ is the fermion propagator of the improved theory
from the point $x$ to the origin.
In order to eliminate  terms of $O(a)$ in the
off-shell correlation functions, used in this study, a further step is
necessary
\cite{wi}:
we have to shift the quark propagator by a local term, which in Fourier space
appears as a constant, to be added to its diagonal components
\beq S^{AB}_{\alpha \beta}(pa) = S^{eff,AB}_{\alpha \beta}(pa)
 + \frac{r}{2}
\delta^{AB}
\delta_{\alpha \beta} \nn \eeq
In the following, the vector current  $V^L_\mu$,  the axial vector current
 $A^L_\mu$, the pseudoscalar density $P$
and the scalar density  $S$ are improved operators,
obtained  by using eq. (\ref{eq:seff}).
\section{Lattice perturbation
theory}
\label{sec:per}
\par
In order to gain some insight into the non-perturbative results
presented in the next section, we have performed the corresponding
one-loop perturbative calculation both in the continuum and
on the lattice  in the Landau gauge. This calculation is an extension of
the one presented in
ref. \cite{newi}.

The standard calculation of the one-loop corrections
corresponds to a computation of  the renormalization constants with a fixed
ultra-violet
cut-off on the loop momenta, $\vert p_i \vert \le \pi/a$,  neglecting terms
of $O(a)$ in the final result. The renormalization constants obtained in
this limit depend on the quark mass $m_q$ and on the external momenta only
through the logarithmically divergent terms, which are related to the anomalous
dimension of the operator. Since, at one loop, these terms are universal, i.e.
they are the same on the lattice and in any  continuum regularization,
the renormalization constants, necessary to relate lattice to continuum
renormalized operators, are independent of the quark mass and external momenta
(they depend however on the lattice spacing and on the continuum
renormalization scale $\mu$). We call this limit ``Standard Perturbation
Theory"
(STP).
The non-perturbative calculation however, is performed  on a lattice of finite
size, and at a fixed value of the lattice spacing and  quark
mass: thus we cannot make  terms of $O(a)$ arbitrarily small. In particular,
since we want to study the $\mu$-dependence of the renormalization constants,
it
is important to know at which values of $\mu$ (and $m_q$),
the perturbative results, obtained on
a finite lattice and with a fixed lattice spacing,
differ from those obtained in SPT. In  order to follow closely the
non-perturbative computation,   the perturbative  results have also been
obtained, on a lattice of size $16^3 \times 32$. In this way,
we keep
 all terms of $O(\alpha_s a)$ or $O(a^2)$, or higher, in the final
result (we call this procedure ``Discrete Perturbation Theory", DPT).   The
loop integrals were
computed  numerically, both in SPT and in DPT.
\par The non-perturbative results have been obtained by tracing suitable
amputated Green functions,  computed between off-shell quark states, in the
Landau gauge, cf. eq.(\ref{eq:proj}). In perturbation theory,  this
procedure differs from the standard one, where one identifies the
 correction from the term  proportional to the
tree-level matrix element of the operator under consideration \cite{buras,ciu}.
 The
difference is due to finite terms  which, in the standard procedure, are
considered as part of the matrix element of the operator, whereas  with the
projectors, are considered as part of the one-loop corrections. At
one-loop order,
these terms are the same in the continuum and on the lattice (in the limit $a
\to 0$) and cancel when one computes the difference between the continuum
and the lattice renormalization constants.
\begin{figure}
\centering
\epsfysize=2.5cm
\epsfxsize=7.0cm
\epsffile{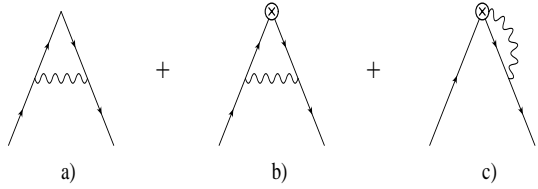}
\caption{Vertex diagrams with the Clover action. The second and
third  diagrams
refer to the contribution from the rotated part of the operator.}
\label{fig:vertex}
\end{figure}
To be more specific, let us
consider the diagrams in fig. \ref{fig:vertex}, computed in
Na\"\i ve Dimensional Regularization (NDR), in the Feynman
gauge
\beq \Lambda_{\gamma_\mu}(\frac{p}{\mu})=
<p|\bar \psi \gamma_\mu \psi|p> =
 \Bigl[ 1+\frac{\alpha_s}{4 \pi} C_F \Bigl(\frac{p^2}{\mu^2}\Bigr)^{-\epsilon}
\Bigl( (\frac{1}{\hat \epsilon}+1) \gamma_\mu-2\frac{\pslash p_\mu}{p^2}
\Bigr) \Bigr]   ,\label{eq:gcont}\eeq
where $C_F=(N^2-1)/2N$, $1/\hat \epsilon=1/\epsilon
-\ln 4 \pi - \gamma_E$
and $p$ is the external momentum. Usually, one takes the
term proportional to $(1/\hat \epsilon +1)$, i.e.
 the coefficient of $\gamma_\mu$,
as the one loop contribution to the bare operator.
 With the projector (see eq.(\ref{eq:proj}))  the last term in
eq. (\ref{eq:gcont}) also gives a contribution, so that the factor now becomes
$(1/\hat \epsilon+1/2)$. In table \ref{tab:cont}, we give the
one-loop corrections obtained, with the projectors and in the
't Hooft and Veltman (HV) dimensional
renormalization scheme \cite{hv}, for the  two-quark operators considered in
this paper.
The NDR regularization differs from the HV one by the
definition of the anti-commuting properties of $\gamma_5$.
For details see for example \cite{ciu}.
 In NDR the corrections to the axial current and pseudoscalar density coincide
with those to the vector and scalar density respectively.
 By writing the gluon propagator as
$G(k)=\left(-\delta_{\mu\nu}k^2+(1-\lambda) k_\mu k_\nu \right)/k^4$,
 we denote as ``Landau" the contribution to the one-loop corrections
proportional to $(1-\lambda)$.
\begin{table}
\begin{center}
\begin{tabular}{||c|c|c||}
\hline
\rm{Operator}
&Feynman&Landau   \\ \hline \hline
$\gamma_\mu$ &$1/\hat \epsilon+ 1/2$ &$-1/\hat
\epsilon-1/2$  \\
\hline
$\gamma_\mu\gamma_5$&$1/\hat \epsilon+5/2$ &$-1/\hat
\epsilon-1/2$     \\ \hline
$\hat 1$  &$4/\hat \epsilon+6$ &$-1/\hat
\epsilon-2$ \\ \hline
$\gamma_5$  &$4/\hat \epsilon+10$ &$-1/\hat
\epsilon-2$ \\ \hline
$\Sigma$  &$-1/\hat \epsilon-1$ &$1/\hat
\epsilon+1$ \\  \hline
\end{tabular}
\caption[]{ \it{The results of the calculation of the one-loop diagram of
fig. \ref{fig:vertex} for different two-quark operators, in the HV scheme, are
given. We also give the result of the calculation of the self-energy
diagram of fig. \ref{fig:self}.
 The mass of the quark has been taken to be
zero, the inverse quark propagator is written as $i \pslash \Sigma(p^2)$
and a factor $(\alpha_s/4 \pi) C_F \Bigl(p^2/\mu^2\Bigr)^{-\epsilon}$
 is omitted.}}
\label{tab:cont}
\end{center}
\end{table}
\begin{figure}
\centering
\epsfysize=2.5cm
\epsfxsize=7.0cm
\epsffile{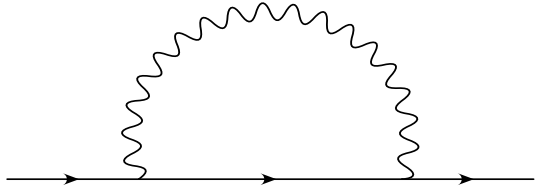}
\caption{Self energy diagram.}
\label{fig:self}
\end{figure}
\par
With the $\overline{{\rm MS}}$ procedure, after the subtraction of the pole,
the
irreducible vertex of fig. \ref{fig:vertex} has the form\footnote{$
\Gamma_O(p/\mu)$ is obtained from $\Lambda_O(p/\mu)$ using a projector, as
described in eq. (\ref{eq:proj}).}
 \beq \Gamma_O(p/\mu)=
  1+\frac{\alpha_s}{4 \pi} C_F
\Bigl(-\gamma^{(O)} \ln (p^2/\mu^2)+C_O^{\overline{{\rm MS}}}
 \Bigr). \label{eq:gms}\eeq
{}From table \ref{tab:cont},  one  finds $C_{\gamma_\mu}^{\overline{{\rm
MS}}} =1/2$, $C_{\hat 1}^{\overline{{\rm MS}}}
=6$, etc., in the Feynman gauge and
$C_{\gamma_\mu}^{\overline{{\rm MS}}} =-1/2$, $C_{\hat 1}^{\overline{{\rm MS}}}
=-2$, etc., for the Landau (longitudinal) terms.
\par On the lattice, one gets a
similar expression \beq \Gamma_O(pa)=
  1+\frac{\alpha_s}{4 \pi} C_F
\Bigl(-\gamma^{(O)} \ln (p^2 a^2)+C_O^{{\rm Latt}}
 \Bigr). \label{eq:gl}\eeq
The values of $C_O^{{\rm Latt}}$, for the Feynman and Landau  corrections are
given in  table \ref{tab:latt}.
\begin{table}
\begin{center}
\begin{tabular}{||c|c|c|c||}
\hline
\rm{Operator}
&Feynman&Rotated Feynman&Landau   \\ \hline \hline
$\gamma_\mu$ &$6.62$ &$-3.52$ &$-4.29$ \\
\hline
$\gamma_\mu\gamma_5$&$5.09$ &
$-11.68$ &$-4.29$ \\ \hline
$\hat 1$ &$16.11$&$-12.00$&$-5.79$ \\ \hline
$\gamma_5$  &$19.18$ &$4.31$ &$-5.79$\\ \hline
$\Sigma$  &$8.21$ &$--$&$4.79$ \\  \hline
\end{tabular}
\caption[]{ \it{ The values of $C_O^{{\rm Latt}}$
 for different two-quark operators and for the self-energy
($C_\Sigma$), with the Clover
action. The mass of the quark has been taken to be
zero.}}
\label{tab:latt}
\end{center}
\end{table}
In table \ref{tab:latt}, we have denoted by ``Rotated Feynman" the contribution
coming from the rotated part of the two-fermion operators, see eqs.
(\ref{eq:oi}) and (\ref{eq:rotaz}). The Landau contribution to the
rotated part of the operators vanishes. This is required by  the fact that the
renormalization constants which relate lattice to continuum operators are
gauge invariant.
\par Following refs. \cite{newi}--\cite{2ferm}, we write
\beq O(\mu)=\left(1+\frac{\as}{4 \pi} C_F \left(-\gamma^{(0)}\ln (\mu a)^2
+\Delta_O \right) \right) O(a) ,\eeq
where $\Delta_O$ can be computed from tables \ref{tab:cont} and \ref{tab:latt}
\beq \Delta_O=C_O^{\overline{{\rm MS}}}-C_O^{{\rm Latt}} +
C_\Sigma^{\overline{{\rm MS}}}-C_\Sigma^{{\rm Latt}} .\eeq
In the difference between continuum and lattice one-loop corrections,
all  Landau pieces cancel, as expected. The results for
$\Delta_{\gamma_\mu},
\Delta_{\gamma_\mu \gamma_5},\cdots$ all agree with those of ref.
\cite{newi}. \par On a lattice, for a given quark mass and
external momenta, there are terms of order $a$  which can become important
at large values of $m_q$ and $p$. Since with our method  we have to impose
the renormalization conditions at values of the momenta
which are large,
it is important to monitor, at least in perturbation theory, the values
of masses and momenta  at which $O(a)$ effects become important. The Green
function in DPT has a form similar to (\ref{eq:gl})
\beq \Gamma_O(pa)=
  1+\frac{\alpha_s}{4 \pi} C_F
\Bigl(-\gamma^{(O)} \ln (p^2 a^2)+C_O^{{\rm Latt}}(pa,m_q a)
 \Bigr), \label{eq:gld}\eeq
where, however, the ``constant" $C_O^{{\rm Latt}}$ now depends on the terms
of $O(a)$. Thus, in the calculation of the one-loop corrections in DPT,
 all the cut-off dependent terms have been included.
On a finite volume, the propagators of massless particles at zero momentum
must be regulated. The results of Feynman diagrams
depend on the regulator, however this dependence vanishes as the volume
is increased. In the results presented below, we have set the zero-momentum
terms in both the quark and gluon propagators to zero. We observe a small
difference in the results from SPT and DPT, at low values of $\mu^2$, which
is due to this choice. At the value of the quark mass at which we
have performed the simulation, mass corrections are negligible.
\par
The coupling constant  appearing in eqs. (\ref{eq:gl}) and (\ref{eq:gld})
is the bare lattice coupling.
In ref. \cite{lepage}   it
was argued that it is possible to optimize the convergence of the perturbative
series by  reorganizing the  expansion in terms of a
``boosted"  coupling $\alpha_s^V$, defined in terms of a physical quantity.
In the comparison with the non-perturbative results below  we follow
the suggestion of ref. \cite{lepage}, and compute
the perturbative corrections using an effective coupling
\beq \alpha_s^V= \frac{1}{\langle \frac{1}{3}{\rm Tr} U_{plaq}\rangle}
\alpha_s^{LATT} \simeq  1.68\,\, \alpha_s^{LATT} ,\eeq
where $\alpha_s^{LATT}=g_0^2/4 \pi$ and
$\langle \frac{1}{3}{\rm Tr} U_{plaq}\rangle$ is the expectation
value of the plaquette. We call this expansion ``Boosted Discrete
Perturbation Theory" (BDPT). Analogously we denote the  ``Boosted
Standard Perturbation Theory" by BSPT.
\par  The results of the calculation of the
renormalization constants in BDPT will be compared to the non-perturbative
results in the next section.
 \section{Non-perturbative results}
\label{sec:nonper}
%___________________________________________________________________________
\begin{figure}[c]   % produce figure here
    \begin{center}
       \setlength{\unitlength}{1truecm}
       \begin{picture}(6.0,6.0)
          \put(-6.0,-6.2){\special{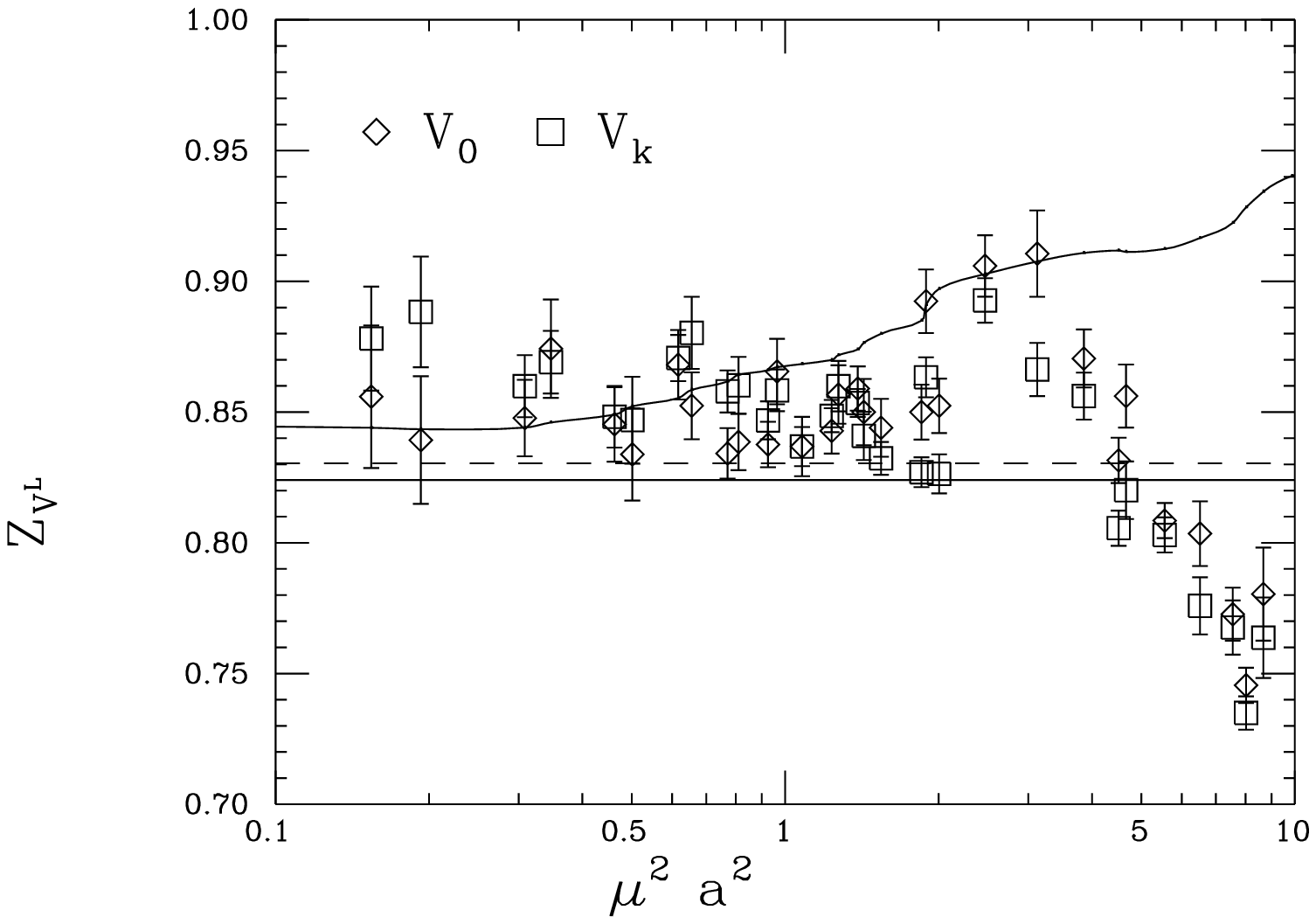}}

       \end{picture}
    \end{center}
       \caption[]{\it{$Z_{V^L}$ as a function of $\mu^2 a^2$.
Since the lattice is not symmetric in space and time, the results
obtained by using the time-component of the local current ($V_0$)
and the space components, averaged in the three directions, ($V_k$)
are shown separately. The dashed line is $Z^{{\rm BSPT}}_{V^L}$ from BSPT,
 and the curve is from BDPT,   on a
volume of size $16^3 \times 32$, see sec. \ref{sec:per},
see sec. \ref{sec:per}.
 The straight
(continous) line is the result obtained using the Ward identities method
\cite{wi,gribov}.}}
    \protect\label{fig:zvl}
\end{figure}
%
%___________________________________________________________________________
\begin{figure}[c]   % produce figure here
    \begin{center}
       \setlength{\unitlength}{1truecm}
       \begin{picture}(6.0,6.0)
          \put(-6.0,-6.2){\special{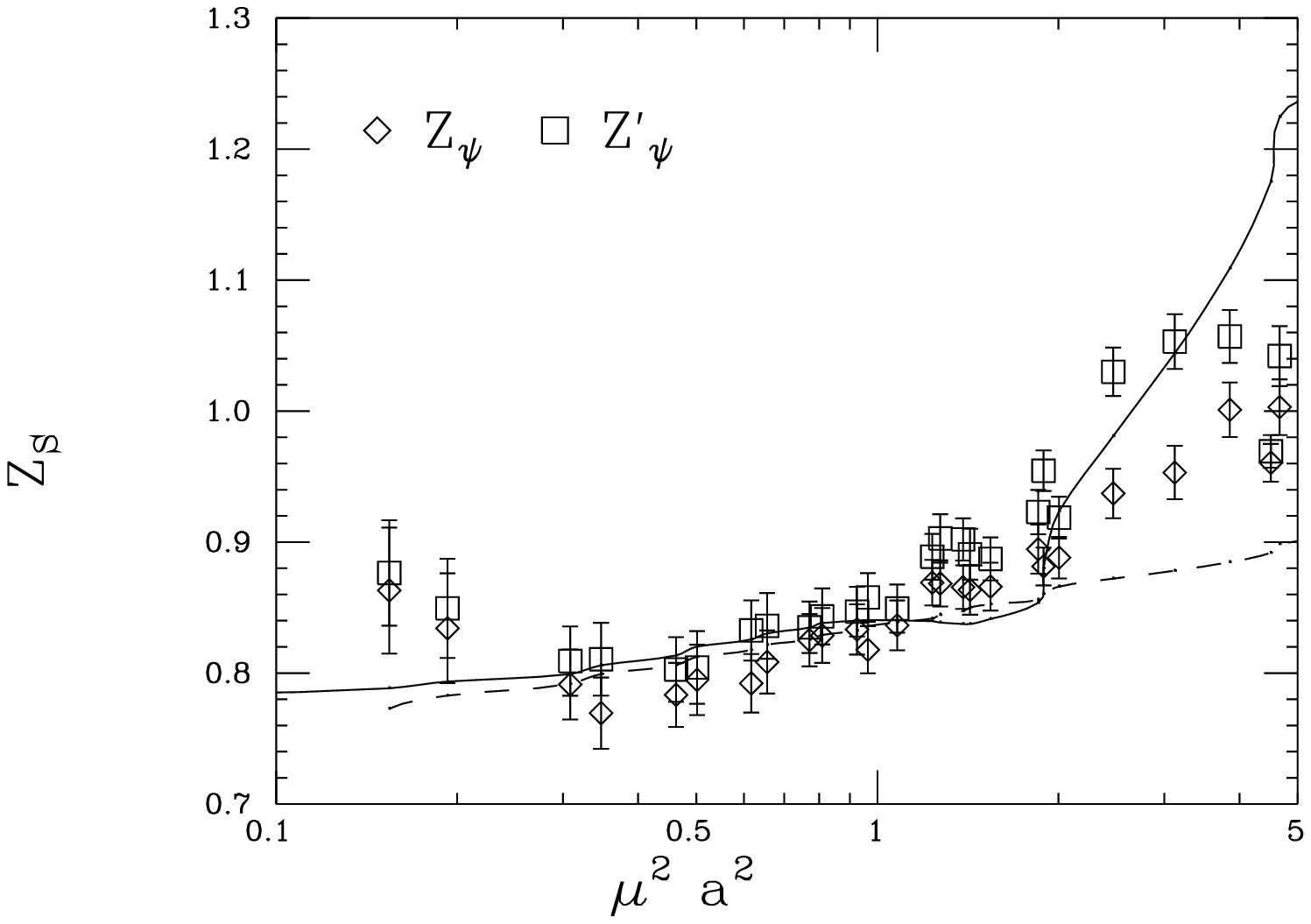}}

       \end{picture}
    \end{center}
       \caption[]{\it{$Z_S$ as a function of $\mu^2 a^2$.
We give the renormalization constant obtained by using the
two possible definitions of the quark wave function renormalization,
denoted by $Z_\psi$, eq. (\ref{eq:zpsil}), and $Z^\prime_\psi$,
 eq. (\ref{eq:zpsi1}).The dashed curve is $Z^{{\rm BSPT}}_S$, from boosted
standard perturbation theory in the infinite volume limit,
 and the full curve is the result obtained in BDPT.}}
    \protect\label{fig:zs}
\end{figure}
%
%___________________________________________________________________________
\begin{figure}[c]   % produce figure here
    \begin{center}
       \setlength{\unitlength}{1truecm}
       \begin{picture}(6.0,6.0)
          \put(-6.0,-6.2){\special{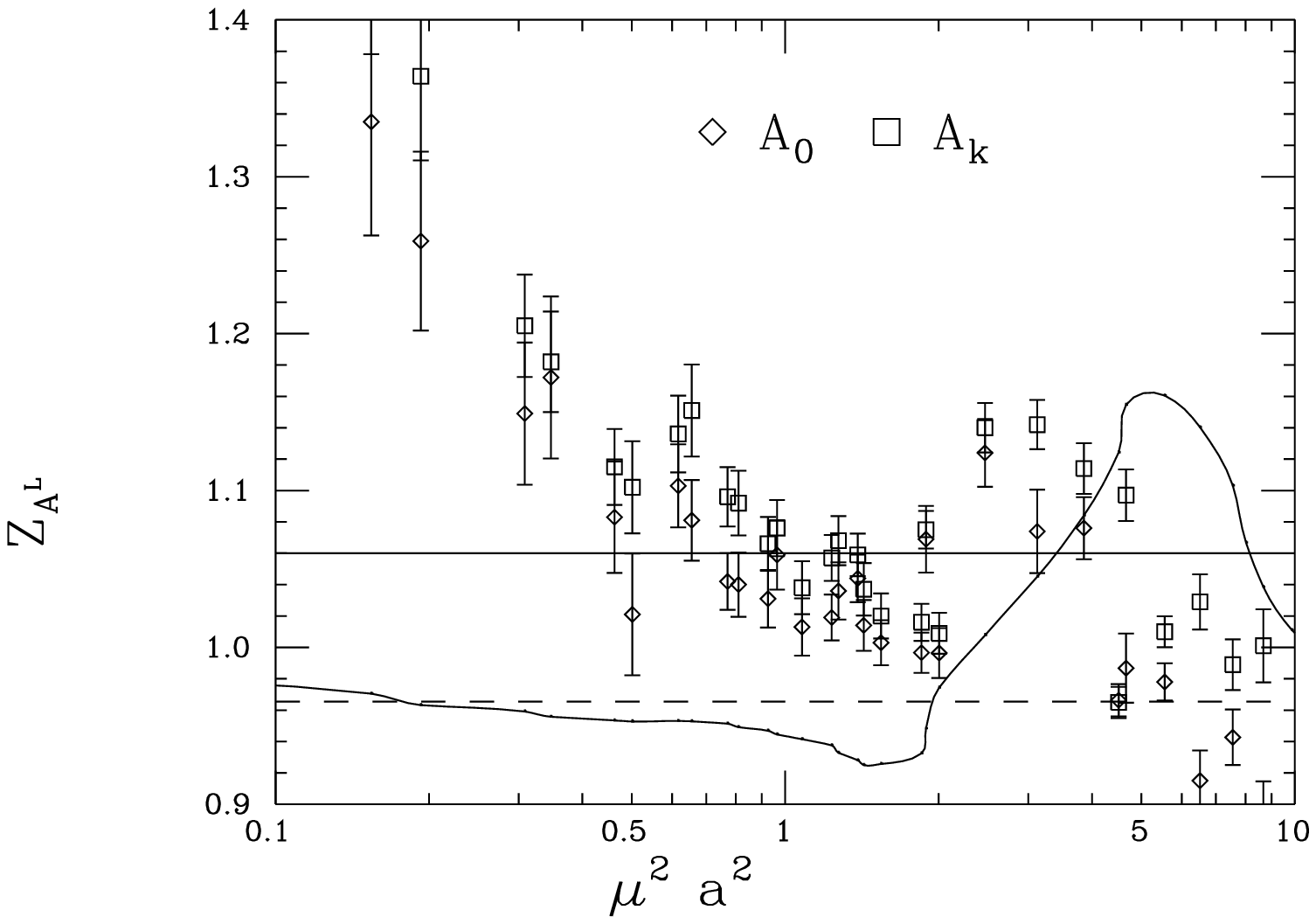}}

       \end{picture}
    \end{center}
       \caption[]{\it{$Z_{A^L}$ as a function of $\mu^2 a^2$.
We give the renormalization constant, obtained by using the
time component $A_0$ and the space components $A_k$, separately.
The dashed line is $Z^{{\rm BSPT}}_A$, from boosted
standard perturbation theory,
 and the full curve is from BDPT. The straight
(continous) line is the result obtained using the Ward identities method
\cite{wi,gribov}.}}
\protect
\label{fig:za}
\end{figure}
%
%___________________________________________________________________________
\begin{figure}[c]   % produce figure here
    \begin{center}
       \setlength{\unitlength}{1truecm}
       \begin{picture}(6.0,6.0)
          \put(-6.0,-6.2){\special{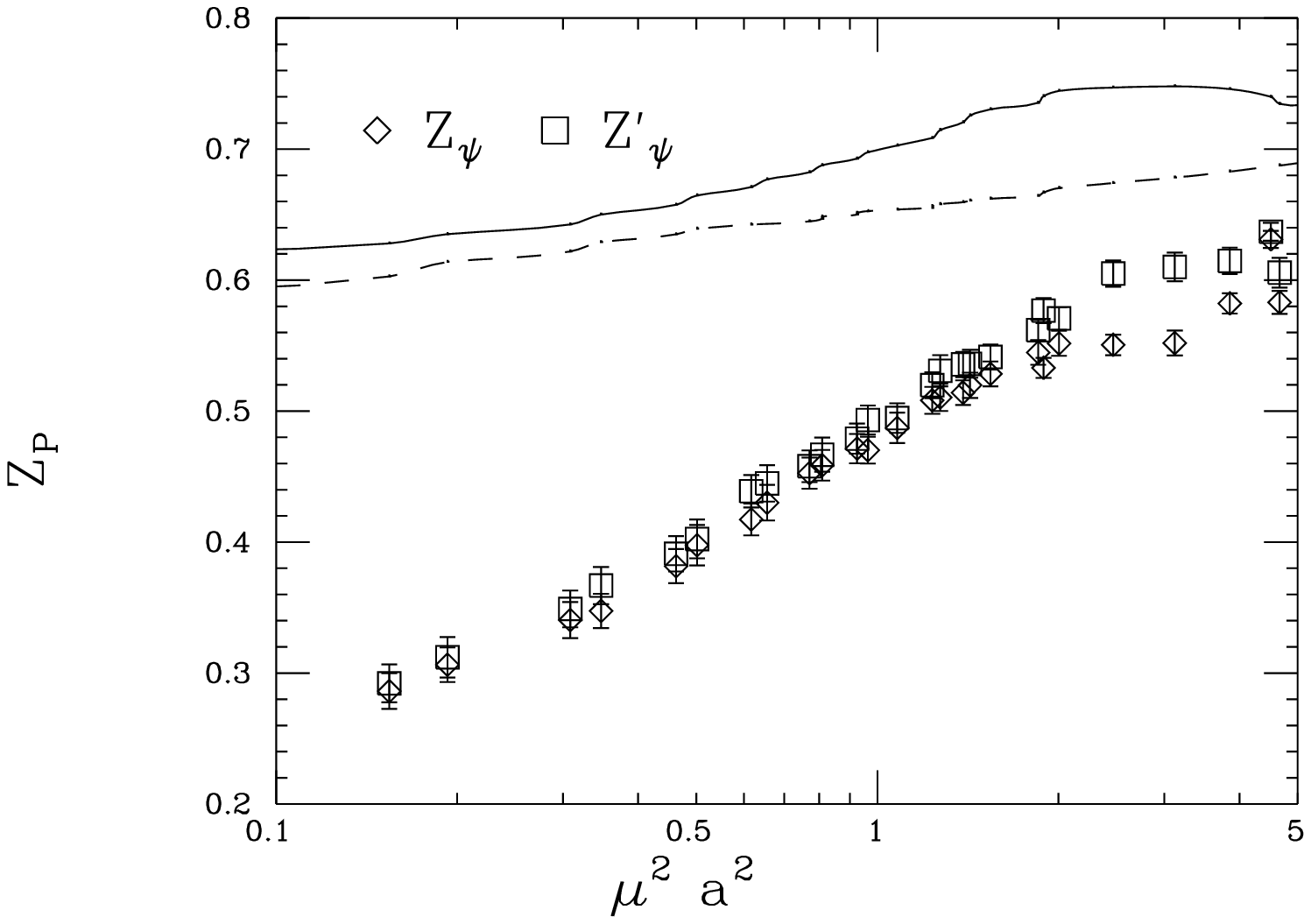}}

       \end{picture}
    \end{center}
       \caption[]{\it{$Z_P$ as a function of $\mu^2 a^2$. We give the
 renormalization constant obtained by using the
two possible definitions of the quark wave function renormalization,
denoted by $Z_\psi$, eq. (\ref{eq:zpsil}), and $Z^\prime_\psi$,
 eq. (\ref{eq:zpsi1}).The dashed curve is $Z^{{\rm BSPT}}_P$, from boosted
standard perturbation theory in the infinite volume limit,
 and the full curve is the result obtained in BDPT.
}}
    \protect\label{fig:zp}
\end{figure}
%
%___________________________________________________________________________
\begin{figure}[c]   % produce figure here
    \begin{center}
       \setlength{\unitlength}{1truecm}
       \begin{picture}(6.0,6.0)
          \put(-6.0,-6.2){\special{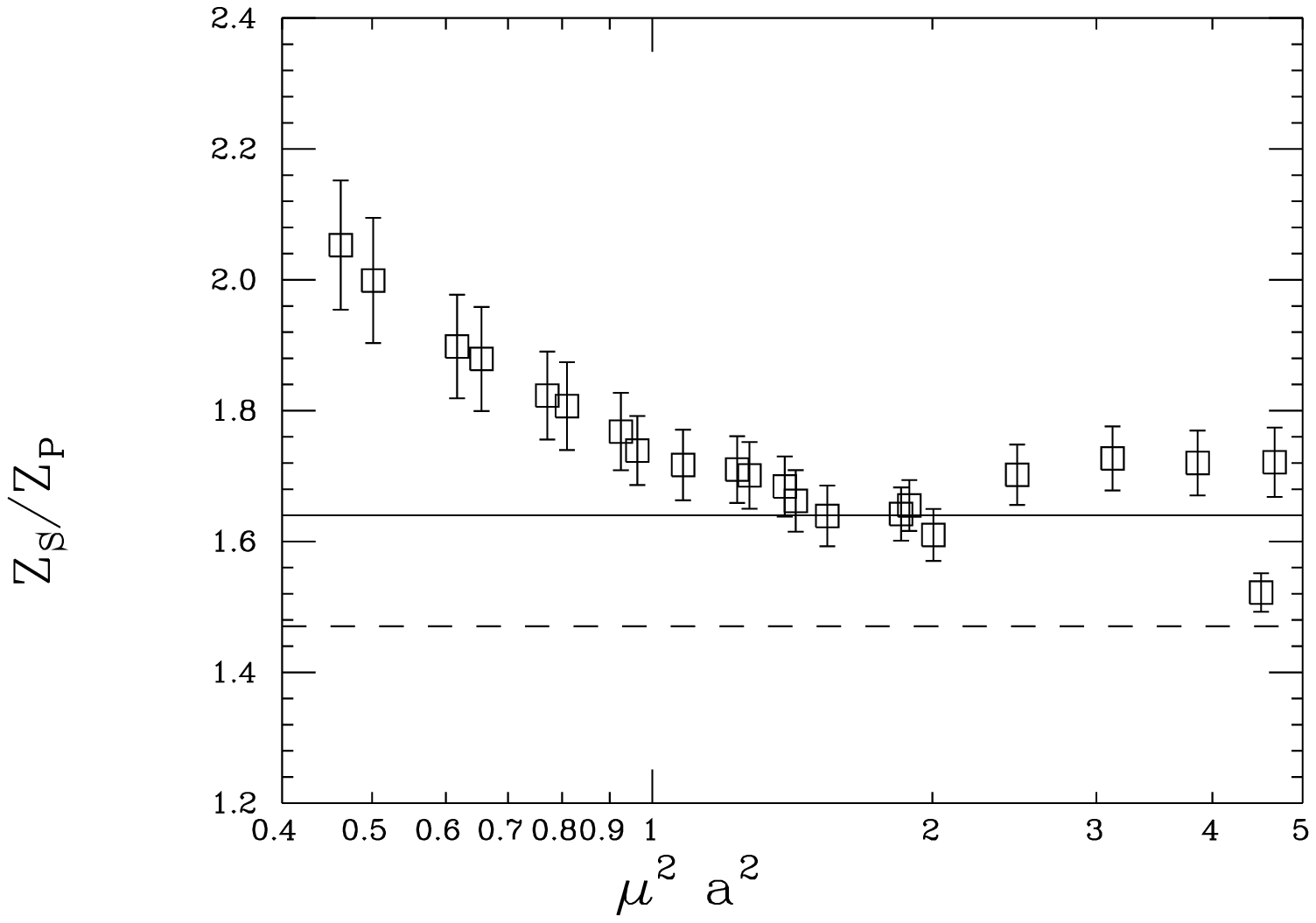}}

       \end{picture}
    \end{center}
       \caption[]{\it{$Z_S/Z_P$ as a function of $\mu^2 a^2$. The dashed line
corresponds to BSPT, the full line comes from the determination of this
ratio, using the Ward identity method \cite{wi}.
}}
    \protect\label{fig:zsszp}
\end{figure}
\par
In this  section, we give the main results of our numerical study of the
non-perturbative renormalization procedure proposed in this paper and a
comparison
of these results with perturbation theory. \par
We have performed a
simulation, by generating $36$ independent gluon field configurations,
on a $16^3 \times 32$ lattice, at $\beta=6.0$. The quark propagators have been
computed for a single value of the quark mass ($am_q \sim 0.07$),
 corresponding to a value of the hopping parameter  $K=0.1425$.
All the Green functions have been computed in the lattice Landau
gauge, which  is obtained by
minimizing the functional \beq
{\rm tr}\Bigl[ \sum_{\mu=1}^{4}\Bigl(U_{\mu}(x)+
U^{\dagger}_{\mu}(x)\Bigr)\Bigr].
\eeq
Possible effects from Gribov copies or spurious solutions \cite{giusti} have
not been considered.
As shown below, for those quantities which can be determined also
in a gauge invariant
way, the non-perturbative results obtained
on quark state  in a fixed gauge  are in
good agreement with  those  obtained using the Ward
identity method  \cite{wi}.
\par {\bf Vector current:}
In fig. \ref{fig:zvl}, the renormalization constant of the local vector current
$Z_{V^L}$, obtained using eqs. (\ref{eq:zpsi1}) and (\ref{eq:gvl}), is given.
As expected, $Z_{V^L}$ is independent of the scale,
whithin  statistical errors, up to large values of $\mu^2$, where distortions
due to lattice discretization become important.
This is a consequence of  the
equivalence, which can be established up to terms of $O(\alpha_s a)$, of the
method used here to determine $Z_{V^L}$, and
of the Ward identity for the local vector current, see sec. \ref{sec:casivari}.
By using the points at $\mu^2 a^2 \sim 1$, corresponding to $\mu \sim 2$ GeV,
we get $Z^{{\rm NP}}_{V^L}=0.84(1)$, to be compared with
$Z^{{\rm WI}}_{V^L}=0.824(2)$  from the Ward identity method, and
$Z^{{\rm BPT}}_{V^L}=0.83$ ($0.86-0.87$ on a $16^3 \times 32$ lattice,
at $\mu^2 a^2 \sim 1$) from BSPT.  The three
methods are in good agreement.
\par {\bf Scalar density.}:
$Z_S$ is an ideal quantity with which to check the validity of
our method for a number of reasons:
it cannot be determined using the Ward identity method, it
is logarithmically divergent and gauge dependent, and
it is not affected by the pole of the Goldstone boson (in contrast to
the pseudoscalar density and the axial current), sec. \ref{sec:idea}.

In fig. \ref{fig:zs}, $Z_S$ is
shown as a function of the renormalization scale.
We report separately the results, obtained by using
$Z_\psi$, eq.(\ref{eq:zpsil}), and $Z^\prime_\psi$,  eq.(\ref{eq:zpsi1}).
Notice that
the points, corresponding to different definitions of $Z_\psi$, are in very
good agreement where $\mu^2 a^2$ is not too large, and
discretization errors
are small. This can also  be seen  by comparing the results in BSPT
(dashed curve) to those in BDPT (continous curve).
We do not expect agreement
with perturbation theory, either at low $\mu^2$, where higher order effects
become very large, or at high $\mu^2$, where lattice distortions are important,
as shown by the continous curve.
The numerical results follow the theoretical expectations and we find good
agreement between the non-perturbative determination of $Z_S$ and the
predictions
from boosted perturbation theory for $0.3 \le \mu^2 a^2 \le 1$
($1.1$ GeV $\le \mu \le 2$ GeV). Notice the $Z_S$ ($Z_P$) is defined here
from the off-shell Green function and depends on the anomalous dimension
 of the scalar (pseudoscalar) density,
and hence  on log($\mu^2 a^2$). Moreover,
the result is gauge dependent, even in the continuum,
 as can be shown by an explicit
calculation in one-loop perturbation theory.
\par {\bf Axial vector current:}
The local axial-vector current shares some of the features of the vector
current.
Its renormalization constant is finite at all orders in perturbation theory
\cite{curci} and can be determined from  Ward identities. However, the
axial-vector current is coupled to the would-be Goldstone boson of $QCD$, which
 can give an important contribution at low $\mu^2$.
$Z_{A^L}$, which like $Z_{V^L}$, should be independent of  $\mu$,
 is
shown as a function of $\mu^2$
in fig. \ref{fig:za}.

 We interpret the strong $\mu$-dependence of
$Z_{A^L}$, at low $\mu^2$, as the non-\-
per\-tur\-ba\-ti\-ve effect of the pseudoscalar state.
Unfortunately, there is no clear sign of the existence of a plateau between the
non-perturbative regime and the large $\mu$ region, where lattice artefacts
become important\footnote{The plateau could eventually appear more clearly at
larger values of $\beta$.}. Without the information coming
from the Ward identity method \cite{wi}, it would be difficult
to determine $Z_{A^L}$ confidently. Nevertheless,
we do observe that, at  $\mu^2 a^2 \sim 1$, just before lattice artefacts
become large in DPT,  the results are close to
the value determined
with the Ward identity, $Z^{{\rm WI}}_A= 1.06(2)$ \cite{wi,gribov}, see
fig. \ref{fig:za}.
As observed in ref. \cite{wi},  perturbation theory gives values of
$Z_{A^L}$ smaller than one, while the Ward identity method
gives values larger than one, for $\beta=6.0$--$6.2$. A value
larger than one is also suggested by our non-perturbative
results.
\par {\bf Pseudoscalar density:} The pseudoscalar density
shares the  main features of the scalar density, with two
important differences. Firstly  it is coupled to the would-be Goldstone
boson, and secondly the one-loop
perturbative corrections, with the Clover action, are quite large.
 It is then not
surprising that the non-perturbative value of the corresponding renormalization
constant $Z_P$ lie well below the perturbative result, as shown in fig.
 \ref{fig:zp}.
 Had we used standard perturbation theory in fig. \ref{fig:zp},
instead of boosted perturbation theory, the discrepancy would
have been even worse. The discrepancy could have been anticipated from the
results of ref. \cite{wi}, where it was shown that the ratio $Z_S/Z_P$,
determined by using the  Ward identity method, was significantly larger than
the result
obtained in boosted perturbation theory. Combining this information
with the agreement between the non-perturbative determination of $Z_S$ and
the result from BPT, one would conclude that the difference is due to
$Z_P$. This may also be due to  the fact that $Z_P$ has
larger one-loop finite corrections, $\sim 35 \%$,
 and that perhaps the  higher order terms, which are not accounted for by
one loop boosted perturbation theory,
are important.

We also present, in fig. \ref{fig:zsszp}, $Z_S/Z_P$ as a function of $\mu^2$.
The ratio $Z_S/Z_P$ has a behaviour similar to $Z_{A^L}$. Some discretization
effects appear to be  smaller than those observed in $Z_S$ and $Z_P$
separately.
The numerical results support, qualitatively, the theoretical
interpretation. In the intermediate range of $\mu$, this ratio stays almost
constant.  At small or large values of $\mu$ the results are
unstable, due to non-perturbative or discretization effects respectively.
 At
$\mu^2 a^2 \sim 1 $, the value of this ratio obtained from our
non-perturbative method  is in reasonable agreement with that determined
with the Ward identity method, $Z_S/Z_P=1.64(5)$ \cite{wi}\footnote
{The value published in ref. \cite{wi}, using half of the
present stastistics,  was $Z_P/Z_S=0.64(2)$,
corresponding to $Z_S/Z_P=1.56(5)$.}.
\section{Conclusion}
\label{sec:concl}
\par
We have proposed a new non-perturbative method to renormalize lattice
composite operators. It can be used in all cases and is particularly
useful when
the Ward identity method is not applicable. This method avoids the need to
perform any calculations using lattice perturbation theory in the
computation of physical quantities from lattice simulations.
 The success
of our proposal is subject to the existence, at current values of
$\beta$, of a window between the non-perturbative region  at low momenta
 and the region of large momenta, where discretization errors become
important. This  limitation is however, common to all methods
(with the exception of the Ward identity method, which can only be  applied
to a few cases, corresponding to finite operators). The window, necessary
to implement  the renormalization programme described here, seems
 to exists already
at $\beta=6.0$ and we expect that the range of momenta, useful
for non-perturbative renormalization, will become larger  as $\beta$
increases. We are planning to apply the approach presented in this paper to the
renormalization of the $\Delta S=2$ four fermion operator (\ref{eq:ds2b}) and
of
the heavy-light axial vector current in the static theory.
%%%%%%%%%%%%%%%%%%%%%%%%%%%%%%%%%%%%%%%%%%%%%%%%%%%%%%%%%%%%%%%%
\section*{Acknowledgements}
\par
We thank S. Petrarca for an early participation to this work and for
many discussions. We also thank E. Gabrielli and the members
of LPTHE
for discussions. G.M. and C.P. thank the theory division at CERN,
where part of this study has been performed. We acknowledge
the
partial support by  M.U.R.S.T. and by the EC contract CHRX-CT92-0051.
C.P. acknowledges the support by the Human Capital and Mobility
Program, contract  ERBCHBICT930887.
CTS acknowledges the Particle Physics and Astronomy Research Council
for its support through the award of a Senior Fellowship.
%%%%%%%%%%%%%%%%%%%%%%%%%%%%%%%%%%%%%%%%%%%%%%%%%%%%%%%%%%%%%%%
\newpage
\begin{figure}
\centering
\epsfysize=2.5cm
\epsfxsize=7.0cm
\epsffile{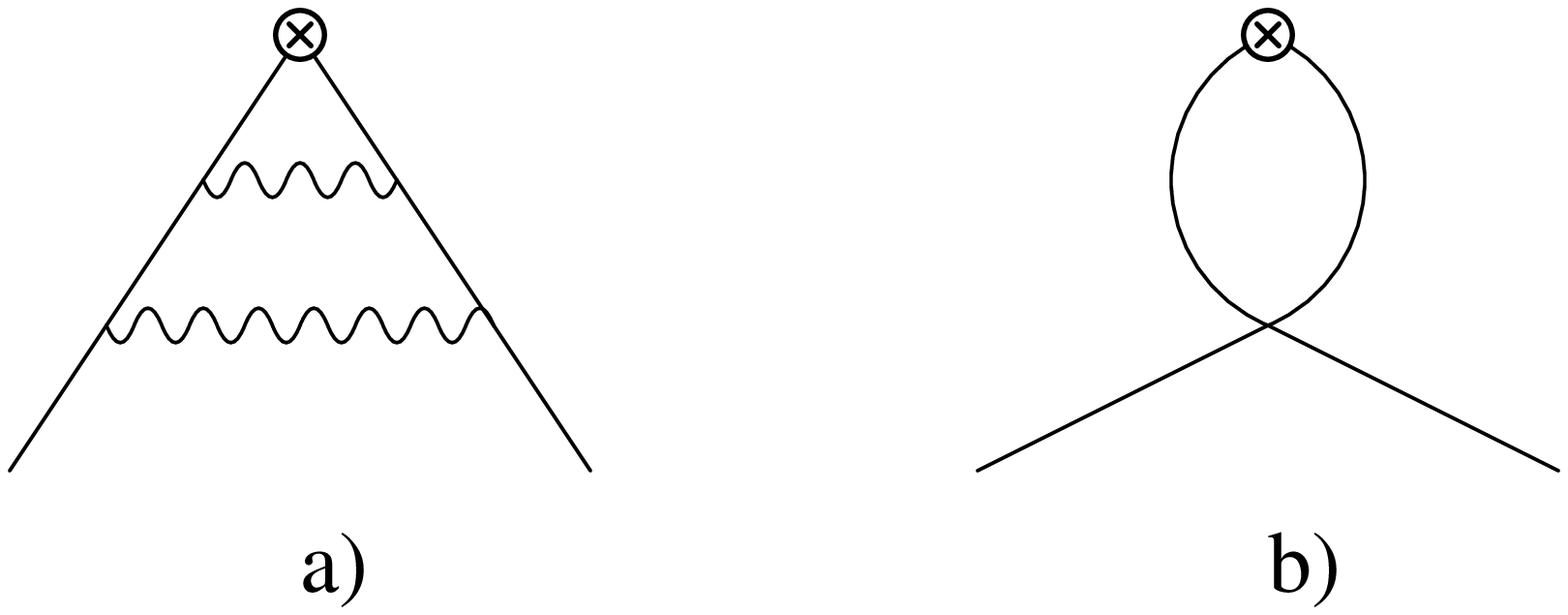}
\caption{Diagrams with standard, a),  and scalar quarks, b).}
\label{fig:appe}
\end{figure}
\section*{Appendix}
In this appendix we demonstrate that the non-perturbative contributions
to forward two-quark matrix elements
are suppressed as inverse powers of $p^2$, where $p$ is the external momentum.
Let us consider the non-amputated, two-fermion Green function
\beq F_O(p)=\int d^4x d^4y e^{-i p \cdot x} \langle \bar q(0) \Gamma
q(x) O_\Gamma(y) \rangle ,\nn \eeq
where $O_\Gamma(y)=\bar q(y) \Gamma q(y)$ and $\Gamma$
is one of the Dirac matrices. It is natural to worry about the infrared
behaviour of  $F_O(p)$, because it contains the insertion
of a zero-momentum operator.  On general
grounds,  we expect infrared divergences to appear in Green functions computed
at exceptional momenta.  In  physical terms, this is due to the fact that
the matrix elements of $O_\Gamma$, at zero momentum, carry
information about the physical mass spectrum, which
 is unaccessible to perturbation theory. In  particular, if
the  operator $O_\Gamma$  is the pseudoscalar
density ($\gamma_5$) and we work in
the chiral limit,  a contribution from  a Goldstone pole, corresponding to the
pion, should appear in $F_O(p)$. The presence of non-perturbative contributions
to $F_O(p)$ is signaled by the appearence of infrared divergences
in the  perturbative expansion. This is what happens in general. For quark
degrees of freedom in the limit of large $p^2$ however, the Operator Product
Expansion  (OPE) guarantees that $F_O(p)$ can be reliably  computed  at all
orders in perturbation theory. The argument goes as follows. In the limit of
large $p^2$, the dominant contribution to $F_O(p)$
comes from  regions of integration (in $x$ and $y$) where
the integrand is singular. This implies that, in this limit, we have to study
the behaviour of
\beq F_O(x,y)=\langle \bar q(0) \Gamma q(x) O_\Gamma(y)  \rangle,\nn \eeq
when $x \sim 0$ and $x \sim y$. In both
cases  of course, we can use the OPE in the following form\footnote{From a
technical point of view, OPE is a weak operator statement and, when $O_\Gamma$
is inserted in a Green function containing other operators, a slightly modified
form would be appropriate. It is easy to convince oneself, however, that the
additionals terms, required in the general case, are not relevant to the
present
discussion.}
\beq q_\alpha (x) O_\Gamma (y) \to A^\Gamma_{\alpha\beta}(x-y)
\Gamma_{\beta\delta} q_\delta (y)+\dots, \nn \eeq
for $x \sim y$, and
\beq \bar q(x) \Gamma q(0) \to B_\Gamma (x^2) \bar q(0) \Gamma q(0) +\dots
,\nn \eeq
for $x \sim 0$.
\par   Since QCD is  asymptotically free, the
  behaviour of $A^\Gamma_{\alpha\beta}(x-y) $ and $B_\Gamma (x)$ can be
computed
in perturbation theory. The interaction gives rise to logarithmic
corrections to the free field behaviour
\beq  A_\Gamma (x) =\xslash C_\Gamma (x^2) \equiv \xslash \frac{a_\Gamma
\ln^{\alpha_\Gamma}(\mu^2 x^2)+\dots}{x^2} \nn \eeq
\beq B_\Gamma(x^2)= b_\Gamma \ln^{\beta_\Gamma}(\mu^2 x^2)+\dots \nn \eeq
Using the above equations, in the limit $p^2 \to \infty$, we find
\beq \int d^4x d^4y e^{-i p \cdot x} \left( \langle \bar q(0)\Gamma  A_\Gamma
(x-y) \Gamma
q(y) \rangle + B_\Gamma(x^2) \langle \bar q(0) \Gamma q(0) O_\Gamma(y) \rangle
\right) = \nn \eeq \beq -i p_\mu \bar C_\Gamma (p^2) \bar S^\mu_\Gamma (p)+
\bar B_\Gamma (p^2) \bar \Delta_\Gamma (0) ,\nn \eeq
where \beq \bar S^\mu_\Gamma (p)= \int d^4y e^{-i p \cdot y}\langle \bar q(0)
\Gamma
\gamma^\mu \Gamma q(y) \rangle
\,\,\,\,\,\,\,\,\,\,\,\,\,
 \bar \Delta_\Gamma (0)= \int d^4y \langle \bar q(0) \Gamma q(0)
O_\Gamma(y) \rangle . \nn \eeq
The term $-i p_\mu \bar C_\Gamma (p^2) \bar S^\mu_\Gamma (p)$ is
computable in perturbation theory. On
  dimensional ground, for $p^2 \to \infty$, it behaves as
\beq -i p_\mu \bar C_\Gamma (p^2) \bar S^\mu_\Gamma (p) \to
c_\Gamma \frac{\ln^{\gamma_\Gamma}(p^2/\mu^2)}{p^2} . \nn \eeq
 As for $\bar B_\Gamma (p^2)$, we find
\beq   \bar B_\Gamma (p^2) \to
d_\Gamma \frac{\ln^{\delta_\Gamma}(p^2/\mu^2)}{p^4} . \nn \eeq
$\bar B_\Gamma (p^2)$   is, however, multiplied by $\bar \Delta_\Gamma (0)$,
 which is a  non-perturbative quantity. We have shown  however,
that, in the large   $p^2$  limit, the perturbative contribution dominates by
one  power of $p^2$ over the non-perturbative one. This  is the
  reason why, even at an exceptional external  momentum,
 $F_O(p)$ is infrared safe  in perturbation theory, when $p^2$ is
  large. It is straightforward to trace the suppression
of the infrared divergence, by analyzing the corresponing Feynman
diagrams.
For example, the infrared divergent part of the
diagram in  fig. \ref{fig:appe}-a is suppressed by one power of $p^2$,
due to the momentum flow through the lower gluon line.
 \par It is interesting to see what would happen in a
(fictitious)
theory, with quarks  represented  by scalar fields. In this case
the infrared divergence is not supressed, as can be seen from
the diagram in fig. \ref{fig:appe}-b.  The general argument goes as follows
\beq  F(p^2) =\int d^4x d^4y e^{-i p \cdot x} \langle \bar \phi (0) \phi(x)
J(y) \rangle, \nn \eeq  with \beq J(y)= \bar \phi (y) \phi(y). \nn \eeq
 When $x \sim 0$ and $x \sim y$, the OPE  gives
\beq  \phi (x) J(y) \to A\left((x-y)^2\right) \phi (y) + \dots
\,\,\,\,\,\,\,\,\,\,\,\,\,
 \bar \phi (x) \phi (0) \to B(x^2) \bar \phi (0) \phi (0) + \dots, \nn \eeq
  with
\beq  A (x^2) =  \frac{a
\ln^{\alpha}(\mu^2 x^2)+\dots}{x^2}
\,\,\,\,\,\,\,\,\,\,\,\,\,
 B(x^2)= b \ln^{\beta}(\mu^2 x^2)+\dots, \nn \eeq
and we would get
\beq F(p^2) \to \bar A(p^2) \bar S(p^2) + \bar B(p^2) \bar \Delta(0) \nn \eeq
with
\beq \bar S (p^2)= \int d^4y e^{-i p \cdot y}\langle \bar \phi(0)
 \phi(y) \rangle
\,\,\,\,\,\,\,\,\,\,\,\,\,
 \bar \Delta (0)= \int d^4y \langle \bar \phi(0)  \phi(0)
J(y) \rangle . \nn \eeq
One now obtains
\beq  \bar A(p^2)  \bar S (p^2) \to
c \frac{\ln^{\gamma}(p^2/\mu^2)}{p^4}  \nn \eeq
and
\beq   \bar B (p^2) \to
d \frac{\ln^{\delta}(p^2/\mu^2)}{p^4} . \nn \eeq
The latter equations imply that non-perturbative and perturbative contributions
have the same $p^2$ dependence. Thus, we expect the appearance of infrared
divergences in $F(p^2)$.

\end{document}